\documentclass{PoS}
\usepackage{graphicx}
\usepackage{epsfig}
\usepackage{amsmath}

\newcommand{\be}{\begin{equation}}
\newcommand{\ee}{\end{equation}}
\newcommand{\bea}{\begin{eqnarray}}
\newcommand{\eea}{\end{eqnarray}}
\newcommand{\beq}{\begin{eqnarray}}
\newcommand{\eeq}{\end{eqnarray}}

\title{The N to $\Delta$ axial transition form factors in quenched and
unquenched QCD }
\ShortTitle{N to $\Delta$ axial form factors}
\author{Constantia Alexandrou\\
Department of Physics, University of Cyprus, CY-1678 Nicosia, Cyprus\\
E-mail:\email{alexand@ucy.ac.cy}}
\author{\speaker{Theodoros Leontiou}\\
Department of Physics, University of Cyprus, CY-1678 Nicosia, Cyprus\\
E-mail:\email{t.leontiou@ucy.ac.cy}}
\author{John W. Negele\\
        Center for Theoretical Physics, Laboratory for
 Nuclear Science and Department of Physics, Massachusetts Institute of
Technology, Cambridge, Massachusetts 02139, U.S.A.\\
        E-mail: \email{negele@mitlns.mit.edu}}
\author{Antonios Tsapalis\\
        University of Athens, Institute of Accelerating Systems
    and Applications, Athens, Greece \\
        E-mail: \email{a.tsapalis@iasa.gr}}
\abstract{

The four N to $\Delta$ axial transition form factors are evaluated
using quenched QCD, using two flavors of dynamical Wilson fermions
and using domain wall valence fermions on three-flavor MILC
configurations for pion masses down to 360 MeV. We provide a
prediction for
 the  parity violating asymmetry as  a function
of $Q^2$ and examine the validity of the non-diagonal
Goldberger-Treiman relation. }

\FullConference{XXIV International Symposium on Lattice Field Theory\\
         July 23-28 2006\\
         Tucson Arizona, US}

\begin{document}

\newcommand{\twopt}[5]{\langle G_{#1}^{#2}(#3;\mathbf{#4};\Gamma_{#5})\rangle}
\newcommand{\threept}[7]{\langle G_{#1}^{#2}(#3,#4;\mathbf{#5},\mathbf{#6};\Gamma_{#7})\rangle}

\section{Introduction}
In this work we present the first lattice QCD evaluation of
 the N to $\Delta$ axial form factors.
A determination of these form factors provides an important
input for the G0 experiment, which is under way to measure these
form factors at Jefferson Lab~\cite{Wells}. The interest in the
axial N to $\Delta$ transition arises from its purely isovector
nature, which
 probes different
physics from what can be extracted from  the study of strange
isoscalar quark currents. Combining results from the
electromagnetic N to $\Delta$ transition we evaluate the dominant
contribution to  the parity violating asymmetry as determined by
the ratio $C_5^A/C_3^V$. This is the  analog of the $g_A/g_V$
ratio extracted from neutron $\beta$-decay. Furthermore  we
investigate low-energy consequences of chiral symmetry, such as
the non-diagonal Goldberger-Treiman relation.

\section{Computational aspects}
Given that this is a first lattice computation of the axial
transition form factors we test our techniques in the quenched
theory where we can use a large volume and have small statistical
errors. The large spatial size of the  lattice allows
 to both reach small $q^2$ values as well as extract the $q^2$-dependence
 more accurately having access to more  lattice momentum vectors
 over a given range of $q^2$.
Pion cloud contributions
are expected to provide
an important ingredient in the description of the
properties of the nucleon system.
 In this work  the light quark regime is
studied with pion masses in the range
of about $(690-360)$~MeV using two degenerate flavors of dynamical Wilson
configurations~\cite{newSESAM,Carsten}
 and  in a hybrid scheme which uses  MILC configurations
generated with staggered sea
quarks~\cite{MILC}   and domain wall valence quarks that preserve chiral
symmetry on the lattice.
 An agreement between  the results from these two
different lattice fermion formulations provides a non-trivial
check of lattice artifacts. In particular finite lattice spacing, $a$, effects
are different: both the quenched and unquenched Wilson fermions
have discretization errors of  ${\cal O}(a)$,
while both Asqtad and domain wall actions have discretization
errors of  ${\cal O}(a^2)$.
Furthermore domain wall fermions preserve chirality,
in contrast to
Wilson fermions.
The hybrid  calculation
is computationally the most demanding. The light quark domain wall
masses are tuned to
reproduce the mass of the Goldstone pion of the staggered sea. Throughout
this work the
 bare quark masses for the domain wall fermions, the size of
the fifth dimension and the renormalization
factors $Z_A$ for the four-dimensional axial vector current are taken
from Ref.~\cite{gaxial}.
In all cases we use Wuppertal
smearing~\cite{Wuppertal}  for the interpolating
fields at the source and sink. In the unquenched Wilson case
to minimize fluctuations~\cite{nucleonff} we use HYP smearing~\cite{HYP} on
the spatial links entering in the Wuppertal smearing of
the source and the sink whereas
for the hybrid case all gauge links in the fermion action are HYP smeared.
In Table~1 we give
the parameters used in our calculation~\cite{axial}.
 The value of the lattice spacing is determined from the nucleon mass
at the chiral limit for the case of Wilson fermions whereas for the hybrid
calculation we take the value determined from heavy quark
spectroscopy~\cite{MILCa}.
\begin{table}[h]\begin{center}
\begin{tabular}{|c|c|c|c|c|}
\hline
\multicolumn{1}{|c|}{no. confs } &
\multicolumn{1}{ c|}{$\kappa$ or $am_l$ } &
\multicolumn{1}{ c|}{$am_\pi$ } &
\multicolumn{1}{ c|}{$aM_N$ } &
\multicolumn{1}{ c|}{$aM_{\Delta}$ }
\\
\hline
\multicolumn{5}{|c|}{Quenched $32^3\times 64$ \hspace*{0.5cm} $a^{-1}=2.14(6)$ GeV}
 \\ \hline
  200            &  0.1554 &  0.263(2) & 0.592(5)   & 0.687(7) \\
  200            &  0.1558 & 0.229(2) &  0.556(6)   & 0.666(8) \\
  200            &  0.1562 & 0.192(2) &  0.518(6)   & 0.646(9)\\
    &  $\kappa_c=$0.1571   & 0.       &  0.439(4)  & 0.598(6)\\
\hline
\multicolumn{5}{|c|}{Unquenched Wilson $24^3\times 40 $~\cite{newSESAM}  \hspace*{0.5cm}
$a^{-1}=2.56(10)$ GeV}
 \\\hline
 185                &  0.1575  & 0.270(3) & 0.580(7) & 0.645(5)\\
 157                &  0.1580  & 0.199(3) & 0.500(10) & 0.581(14) \\
\hline
\multicolumn{5}{|c|}{Unquenched Wilson $24^3\times 32 $~\cite{Carsten}  \hspace*{0.5cm}
$a^{-1}=2.56(10)$ GeV}
\\\hline
 200                &  0.15825 & 0.150(3) & 0.423(7)  & 0.533(8)  \\
                    & $\kappa_c=0.1585$& 0. & 0.366(13)& 0.486(14)\\
\hline
\multicolumn{5}{|c|}{MILC $20^3\times 64 $  \hspace*{0.5cm}
$a^{-1}=1.58$ GeV}
 \\\hline
 150                &  0.03  & 0.373(3) & 0.886(7) & 1.057(14)\\
 150                &  0.02  & 0.306(3) & 0.800(10)&  0.992(16)\\
\hline
\multicolumn{5}{|c|}{MILC $28^3\times 64 $  \hspace*{0.5cm}
$a^{-1}=1.58$ GeV}
\\\hline
 118                &  0.01 & 0.230(3) & 0.751(7)  & 0.988(26)  \\
\hline
\end{tabular}
\end{center}
\vspace*{-0.5cm}\caption{The number of configurations,
 the hopping parameter, $\kappa$, for the
case of  Wilson fermions or the mass of the light quarks, $m_l$,
for the case of staggered quarks,
 the pion,  nucleon and $\Delta$ mass in lattice
units.} \label{table:parameters}
\end{table}

\section{Methodology}
The calculation of the axial form factors makes use of  the
same methodology as the one used in our lattice study
of the electromagnetic N to $\Delta$ transition~\cite{ND,dina}.
The invariant N to $\Delta$ weak matrix element
can be expressed in terms of four transition form factors as
\beq
&&<\Delta(p^{\prime},s^\prime)|A^3_{\mu}|N(p,s)> = i\sqrt{\frac{2}{3}}
\left(\frac{M_\Delta M_N}{E_\Delta({\bf p}^\prime) E_N({\bf p})}\right)^{1/2}\bar{u}^\lambda(p^\prime,s^\prime)
\nonumber \\
&&\hspace*{-1cm}\biggl[\left (\frac{C^A_3(q^2)}{M_N}\gamma^\nu + \frac{C^A_4(q^2)}{M^2_N}p{^{\prime \nu}}\right)
\left(g_{\lambda\mu}g_{\rho\nu}-g_{\lambda\rho}g_{\mu\nu}\right)q^\rho
+C^A_5(q^2) g_{\lambda\mu} +\frac{C^A_6(q^2)}{M^2_N} q_\lambda q_\mu \biggr]u(p,s)
\label{matrix element}
\eeq
where $q_\mu=p{^\prime}_\mu-p_\mu$ is the momentum transfer and
 $A^3_\mu(x)=  \bar{\psi}(x)\gamma_\mu \gamma_5 \frac{\tau^{3}}{2} \psi(x)$
is the isovector part of the axial current ($\tau^3$ being
the third Pauli matrix).
In order to evaluate this matrix element on the lattice we compute the three
point function  \(\threept{\sigma}{\Delta j^\mu
N}{t_2}{t_1}{p'}{p}{}\).
We  eliminate the exponential decay in time and
the overlaps of the interpolating fields with the physical states
by forming an appropriate  ratio,
$R_\sigma (t_2, t_1; {\bf p}^\prime,{\bf p};\Gamma;\mu)$,
of three-point and two-point functions
given by
\begin{eqnarray}
R_\sigma&=&\frac{\langle G^{\Delta j^\mu N}_{\sigma} (t_2, t_1 ;  {\bf p}^\prime,{\bf p}  ;\Gamma ) \rangle \;}{\langle G^{\Delta }_{ii} (t_2, {\bf p}^\prime;\Gamma_4 ) \rangle \;}
\biggl [ \frac{ \langle G^{N}(t_2-t_1, {\bf p};\Gamma_4 ) \rangle \;\langle
G^{\Delta }_{ii} (t_1, {\bf p}^\prime;\Gamma_4 ) \rangle \;\langle
G^{\Delta }_{ii} (t_2, {\bf p}^\prime;\Gamma_4 ) \rangle \;}
{\langle G^{\Delta}_{ii} (t_2-t_1, {\bf p}^\prime;\Gamma_4 ) \rangle \;\langle
G^{N} (t_1, {\bf p};\Gamma_4 ) \rangle \;\langle
G^{N} (t_2, {\bf p};\Gamma_4 ) \rangle \;} \biggr ]^{1/2} \nonumber \\
&\>&\overset{t_2-t_1\gg1,t_1\gg1}{\Rightarrow}\Pi_\sigma(\mathbf{p'},\mathbf{p};\Gamma;\mu),
\label{ratio}
\end{eqnarray}
where
$\Gamma_4=\frac{1}{2}\left(\begin{array}{cc} I &0\\0 & 0 \end{array}\right)$
and  $\Gamma_j=\frac{1}{2}\left(\begin{array}{cc} \sigma_j &0\\0 & 0 \end{array}\right)$.
With $t_1$ we denote the time when a photon interacts with
a quark  and with $t_2$, the time when the $\Delta$ is annihilated.
The  ratio given in Eq.~(\ref{ratio})
is constructed so that the two-point functions that enter have the shortest
possible time separation between source and sink. This
  provides an optimal signal to noise ratio.
For large time separations $t_1$ and $t_2$, the ratio
$R_\sigma(t_2,t_1;\mathbf{p'},\mathbf{p};\Gamma;\mu)$ becomes time
independent and yields the transition matrix element of
Eq.~(\ref{matrix element}) up to the renormalization constant
$Z_A$. The latter has been computed non-perturbatively
using the RI-MOM method for quenched~\cite{renorm} and two flavor of
dynamical Wilson fermions~\cite{renorm_dyn}.
 The values obtained in
both cases are all consistent with   $Z_A=0.8$. For domain wall
fermions we use the values given in Ref.~\cite{gaxial}. We use
kinematics where the $\Delta$ is produced at rest and by
$Q^2=-q^2$ we denote the Euclidean momentum transfer squared.

There are various choices for the Rarita-Schwinger spinor index
$\sigma$ and projection matrices $\Gamma$ that yield the four
axial form factors. Each of these choices requires a separate
sequential inversion. As in the case of the  evaluation of the
electromagnetic N to $\Delta$ transition form factors~\cite{ND} we
use optimized $\Delta$ sources in order to maximize the number of
lattice momentum vectors contributing to a given $Q^2$  value.
The optimized $\Delta$ sources turn out to be  the same as those
used in our study of the electromagnetic form factors~\cite{ND}. Namely
we use the combinations
$S_1(\mathbf{q};\mu)=\sum^3_{\sigma=1}\Pi_\sigma(\mathbf{q};\Gamma_4;\mu)$,
$S_2(\mathbf{q};\mu)=\sum^3_{\sigma\ne k=1}\Pi_\sigma(\mathbf{q};\Gamma_k;\mu)$
and $S_3(\mathbf{q};\mu)=\Pi_3(\mathbf{q};\Gamma_3;\mu)-\frac{1}{2}(\Pi_1(\mathbf{q};\Gamma_1;\mu)+\Pi_2(\mathbf{q};\Gamma_2;\mu)) $.
The four axial form factors can be extracted from the following expressions
\beq \nonumber
S_1(\mathbf{q};j)&=&i B\Bigg[-\frac{C^A_3}{2(E_N+M_N)}\bigg\lbrace (E_N+M_N)(E_N-2M_\Delta+M_N)+\left(\sum_{k=1}^3 p^k\right)p^j \bigg\rbrace \\
\nonumber &-&\frac{M_\Delta}{M_N}(E_N-M_\Delta)C^A_4
+M_NC^A_5-\frac{C^A_6}{M_N}p^j\left(\sum_{k=1}^3 p^k\right)\Bigg],~~~j=1,2,3\\ \nonumber
S_1(\mathbf{q};4)&=&B\sum_{k=1}^3 p^k\Bigg[C^A_3+\frac{M_\Delta}{M_N}C^A_4
+\frac{E_N-M_\Delta}{M_N}C^A_6\Bigg],\\ \nonumber
S_2(\mathbf{q};j)&=&i A\Bigg[\frac{3}{2}\left(\sum_{k=1}^3 p^k\right)\left(\delta_{j1}(p^2-p^3)+
\delta_{j2}(p^3-p^1)+\delta_{j3}(p^1-p^2)\right)C^A_3\Bigg],\\
S_3(\mathbf{q};j)&=&i A\Bigg[\frac{9}{4}\left(
\delta_{j1}(p^2p^3)-\delta_{j2}(p^1p^3)\right)C^A_3\Bigg],
\eeq
where $A=\sqrt{2/3}\sqrt{E_N/(E_N+M_N)}/(3E_N M_N)$ and $B=A/(E_N+M_N)$.
The axial form factors can be extracted
by performing an overconstrained analysis as described in
Refs.~\cite{ND,dina}.

\section{Results}


\begin{figure}[h]
 \begin{minipage}[h]{7cm}
{\mbox{\includegraphics[height=8cm,width=7.5cm]{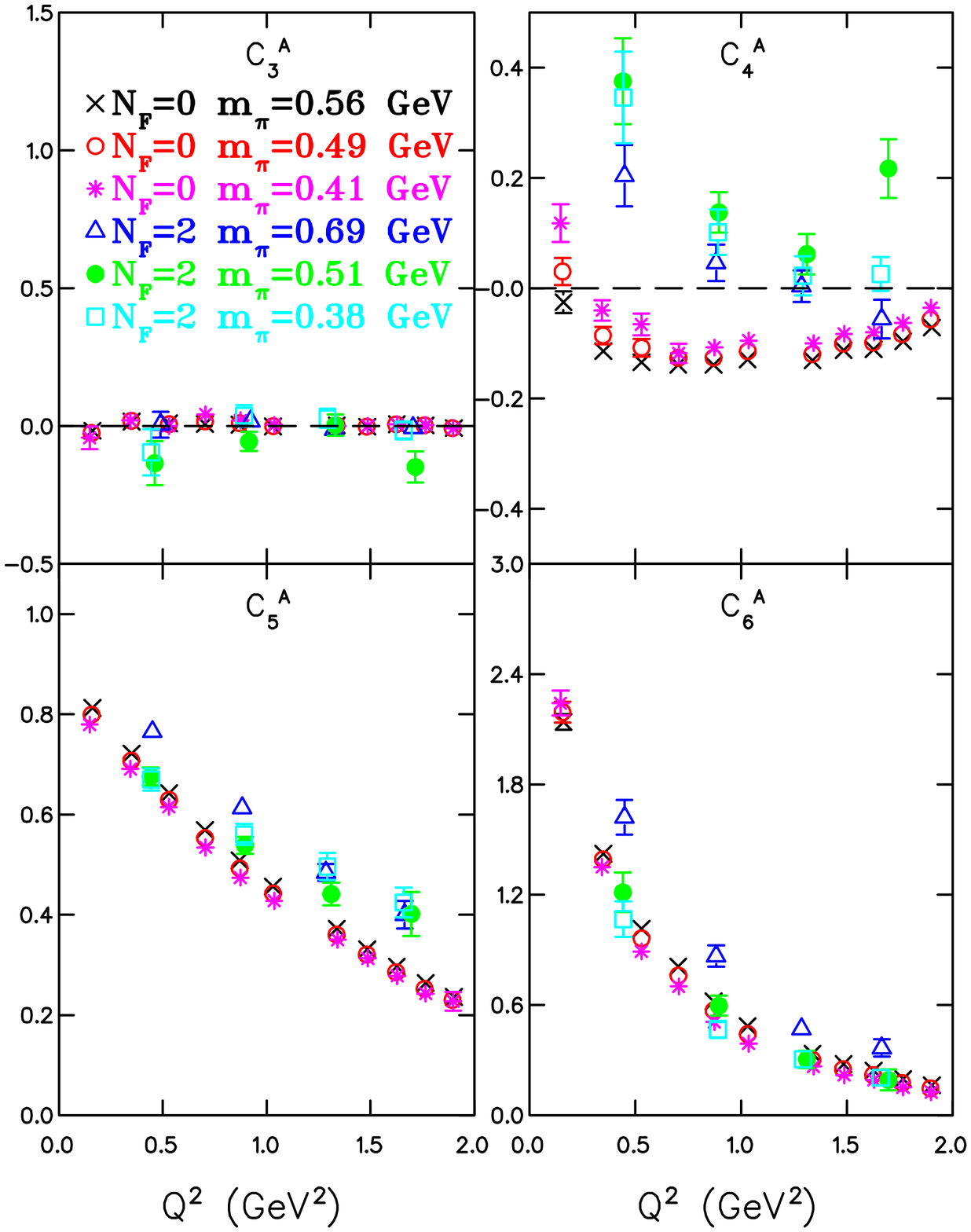}}}
    \end{minipage}
    \hfill
    \begin{minipage}[h]{7cm}\hspace*{-0.5cm}
{\mbox{\includegraphics[height=8cm,width=7.5cm]{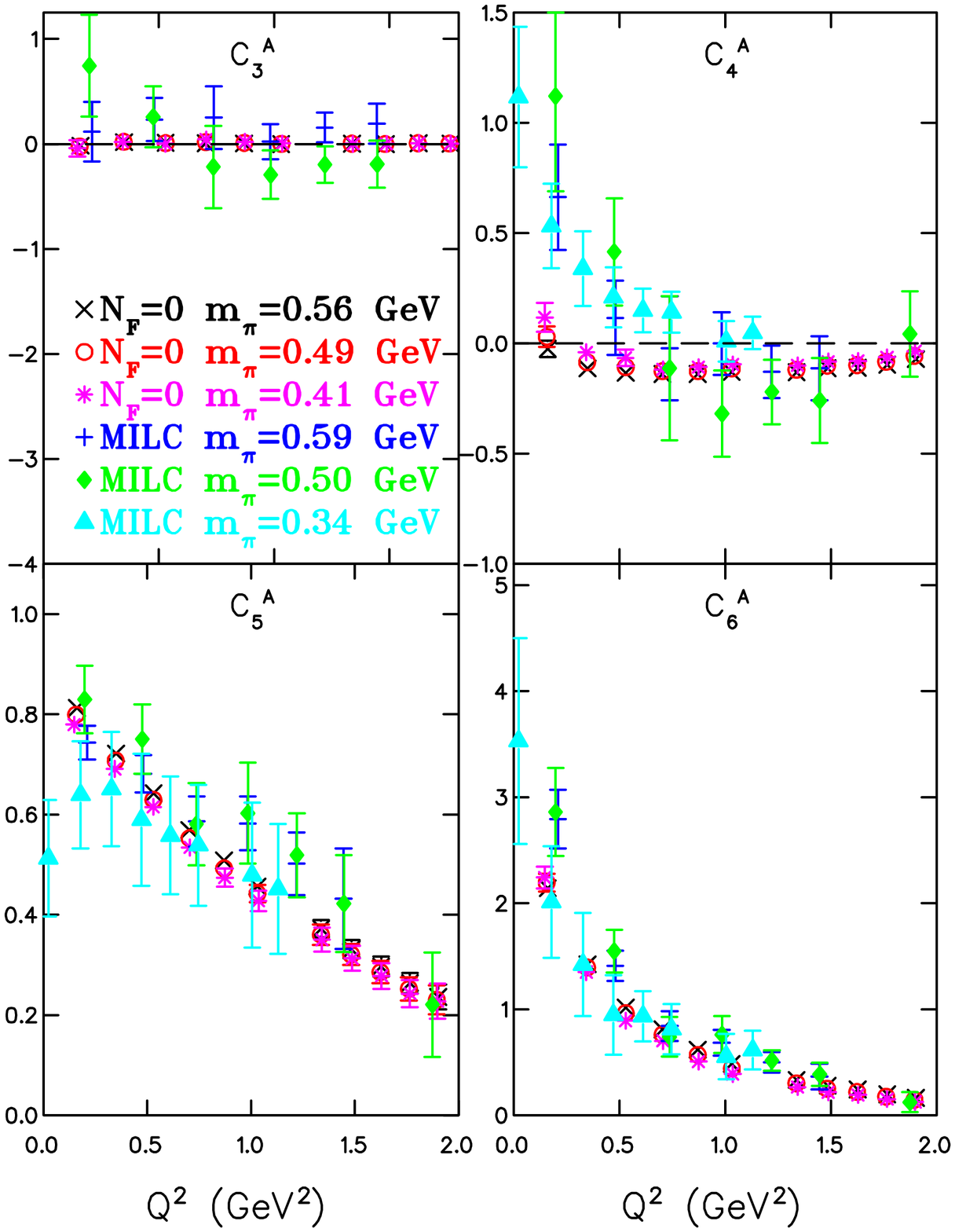}}}
    \end{minipage}
    \caption{\label{fig:CA}
The axial form factors $C^A_3$, $C^A_4$, $C^A_5$ and $C^A_6$
    as a function of $Q^2$.  In all plots
we show  quenched  results,  denoted by $N_f=0$, at
$\kappa=0.1554$ (crosses), at $\kappa=0.1558$ (open circles) and
at $\kappa=0.1562$ (asterisks).
 The graphs on the left hand side also show
unquenched Wilson results,  denoted by $N_f=2$,
 at $\kappa=0.1575$ (triangles),
$\kappa=1580$ (filled circles) and $\kappa=0.15825$ (open squares).
The graphs on the right hand side also show results in the hybrid approach
    at $am_l=0.03$ (crosses), $am_l=0.02$ (filled diamonds) and $am_l=0.01$ (filled triangles).}
\vspace*{-0.4cm}
\end{figure}
In Fig.~\ref{fig:CA}  we
show our lattice results for the four axial form factors for
quenched and unquenched  Wilson fermions and
in the hybrid approach.
 We observe that $C_3^A$ is
consistent with zero and that unquenching effects
are small for the dominant form factors, $C_5^A$ and $C_6^A$.
The form factor $C_4^A$ shows an interesting
behavior: The unquenched results for both dynamical
Wilson and domain wall fermions
show an increase at
low momentum transfers. Such large deviations between
quenched and full QCD results for these relatively heavy
 quark masses  are unusual
making  this  an interesting quantity to study
effects of unquenching.

\begin{figure}
    \begin{minipage}[h]{7cm}
{\mbox{\includegraphics[height=9cm,width=7cm]{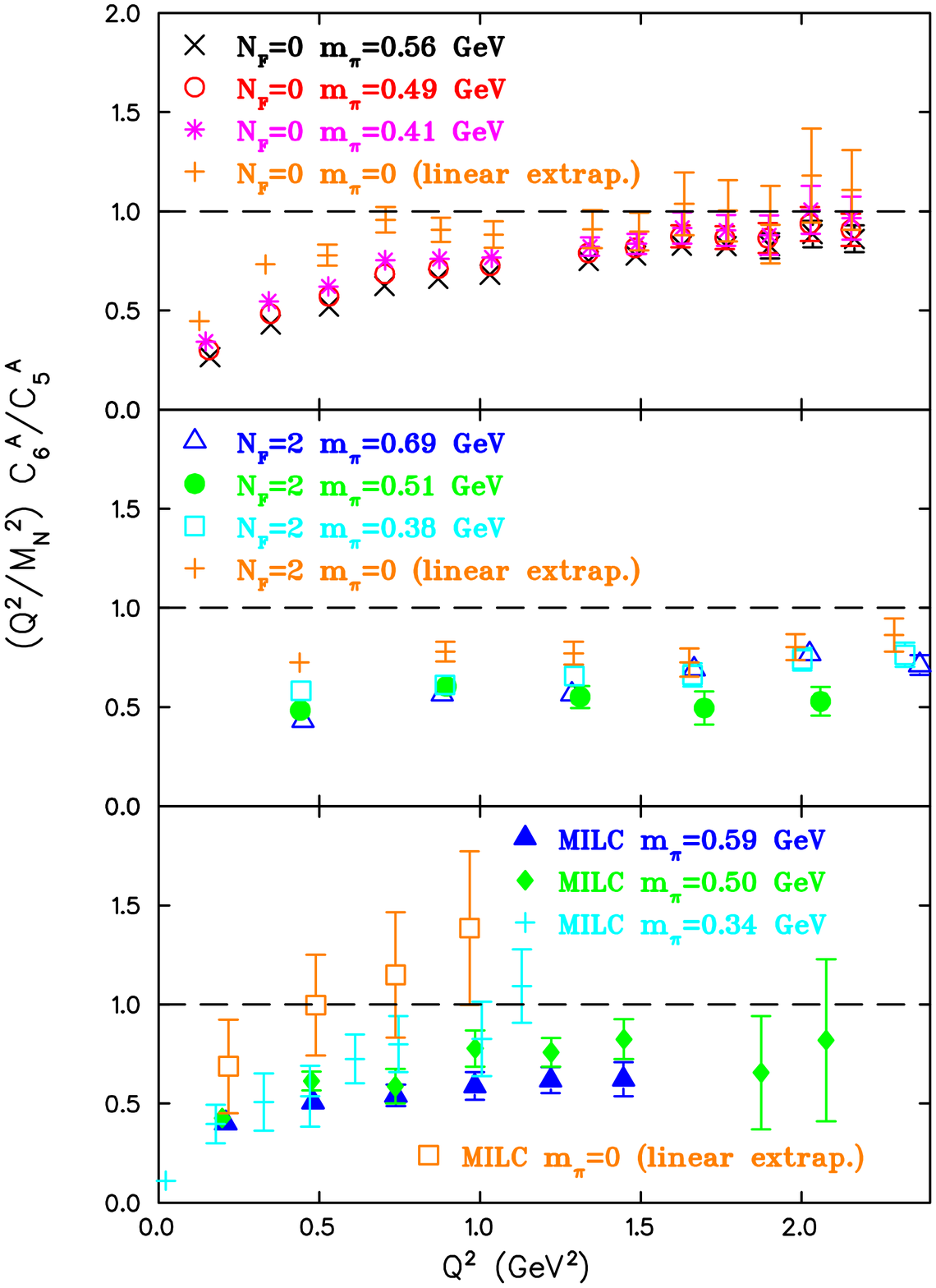}}}
      \caption{The ratio $\left(Q^2/M_N^2\right) C^A_6/C^A_5$
 versus $Q^2$. Top: In the quenched theory,
Middle: For dynamical Wilson fermions,  
       Bottom: In the hybrid approach.
      \label{fig:light quark ratio}}
    \end{minipage}\hfill
    \begin{minipage}[h]{7cm}\vspace*{-1cm}
{\mbox{\includegraphics[height=8cm,width=7.cm]{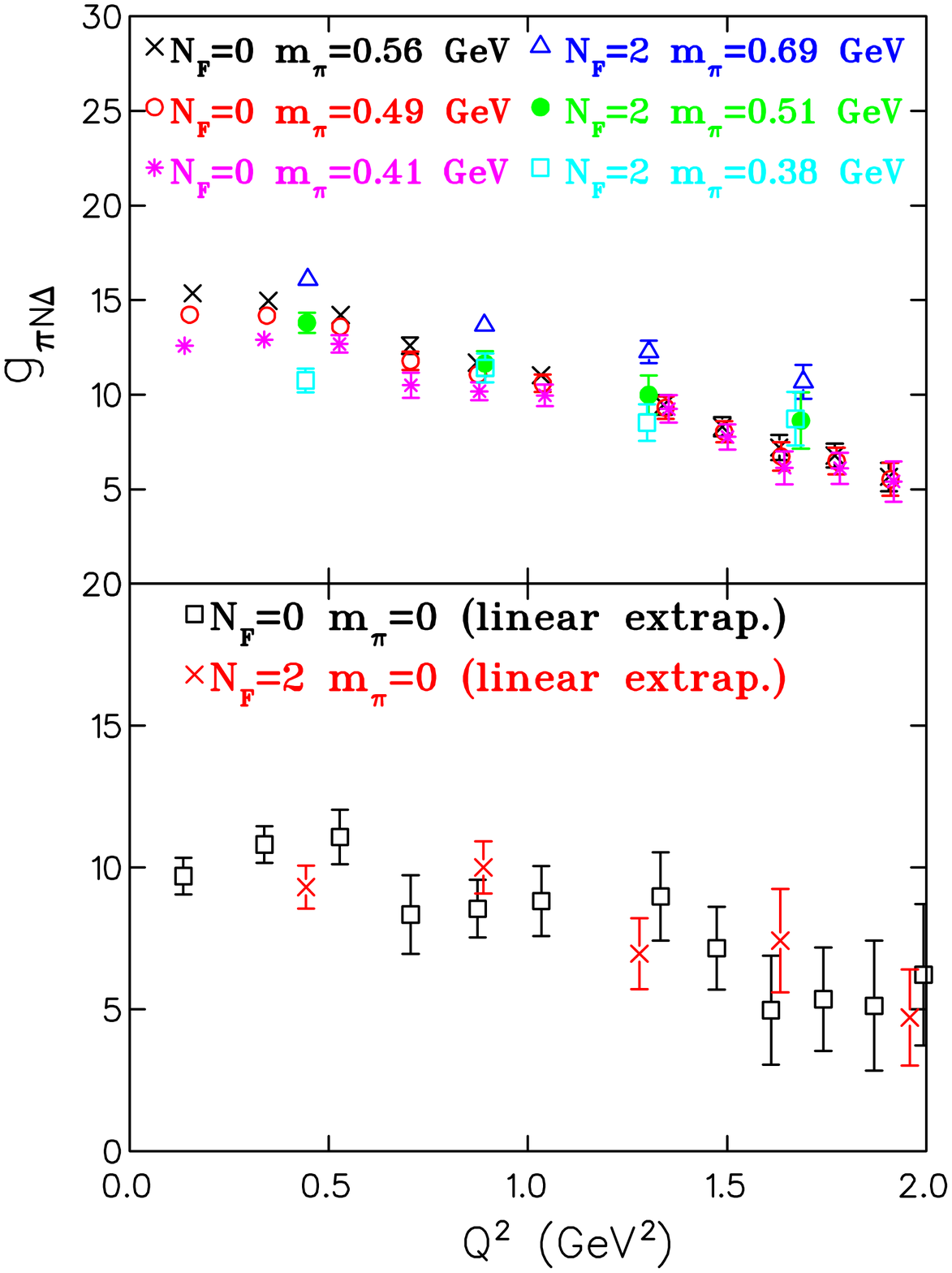}}}
\caption{\label{fig:gpind}$g_{\pi N\Delta}$ as a function of $Q^2$
for quenched and dynamical Wilson fermions.
Top: For our three-$\kappa$ values.
Bottom: In the chiral limit.
}
    \end{minipage}
\vspace*{-0.5cm}
\end{figure}

\begin{figure}[h]
\begin{minipage}[h]{6.5cm}
{\mbox{\includegraphics[height=5cm,width=7.cm]{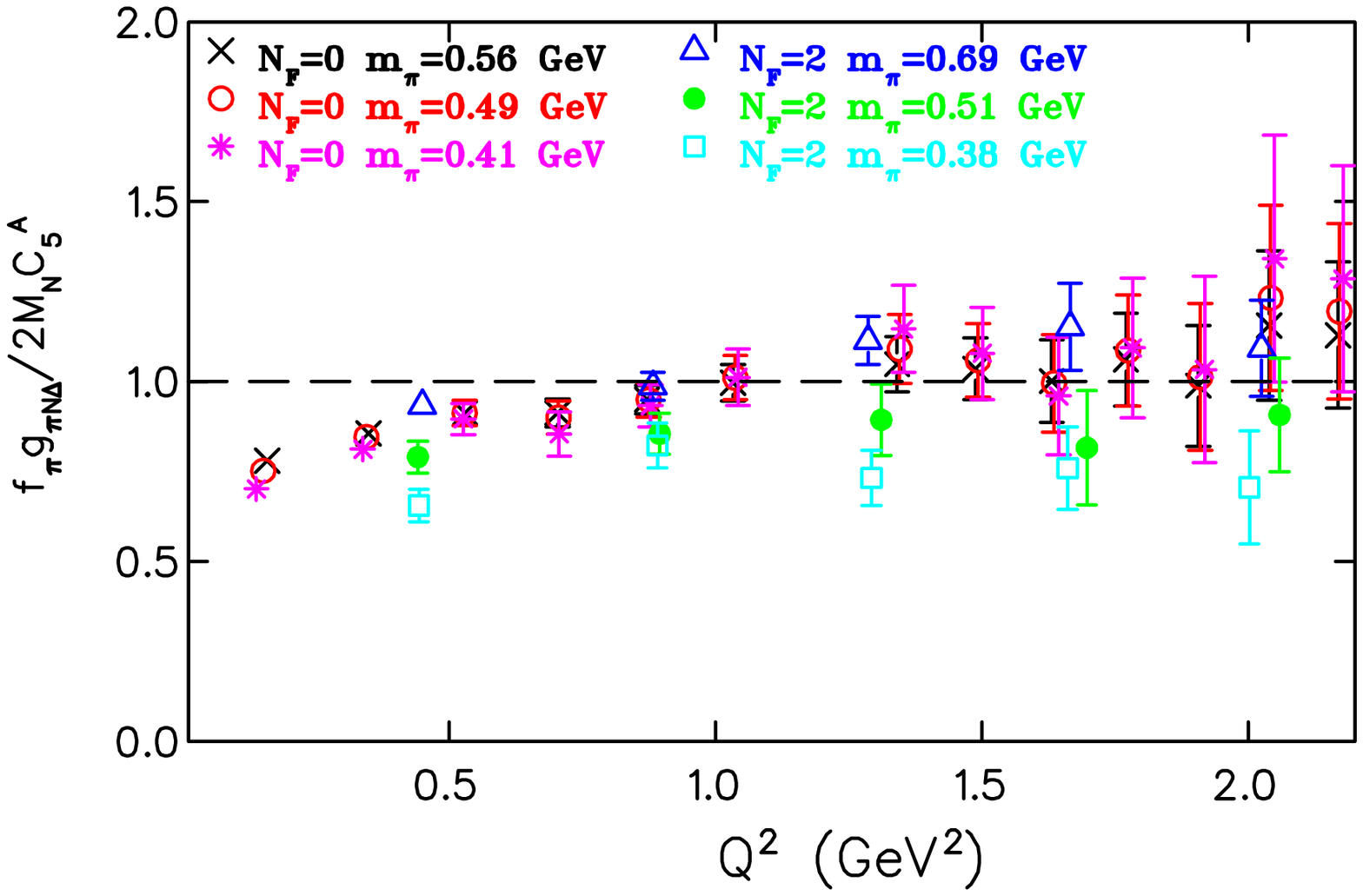}}}
\caption{The ratio $f_\pi g_{\pi N\Delta}/\left(2M_N C^A_5\right)$
 versus $Q^2$ for quenched and dynamical Wilson fermions. 
\label{fig:Goldberger-Treiman}}
\end{minipage}\hfill
\begin{minipage}[h]{7.5cm}
{\mbox{\includegraphics[height=5cm,width=7.cm]{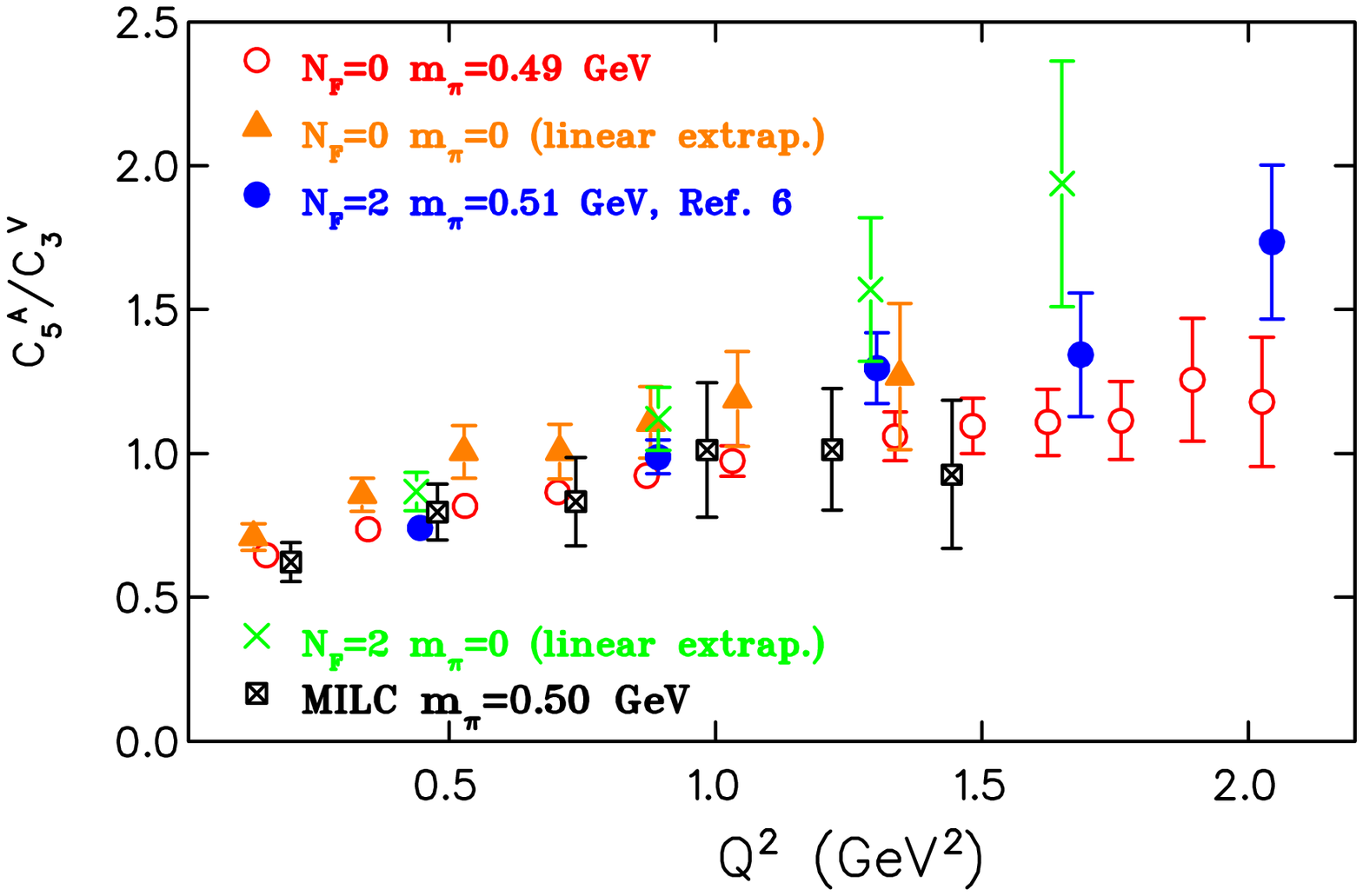}}}
\caption{The ratio $C^A_5/C^V_3$ as a function of $Q^2$ for
quenched QCD, for dynamical Wilson fermions and for the hybrid
scheme.\label{fig:CA5 over CV3}}
\end{minipage}
\vspace*{-0.5cm}
\end{figure}

In the chiral limit, axial current conservation
leads to the relation $C_6^A(Q^2)=M_N^2 C_5^A(Q^2)/Q^2$.
In Fig.~\ref{fig:light quark ratio} we show the ratio
$\left(Q^2/M_N^2\right) C_6^A(Q^2)/C_5^A(Q^2)$
for quenched and unquenched Wilson fermions,
and in the hybrid scheme.
In each case we show results for the available quark masses and
in the chiral limit. The expected value in the chiral limit for
this ratio is one.
For finite quark mass the axial current is not conserved and
for Wilson fermions
chiral symmetry is broken so that deviations from one are expected.
We observe  that this ratio differs from unity at low $Q^2$ but
approaches unity at higher values of $Q^2$.
For the hybrid scheme the ratio is consistent with
unity even at the lowest available $Q^2$,
as expected for chiral fermions.
That such  chiral restoration is seen on the lattice even when using
Wilson fermions demonstrates that lattice methodology correctly encodes
continuum physics.


At finite pion mass partial conservation of axial current
($\partial_\mu A_\mu^a(x)=f_\pi m_\pi^2\pi^a(x)$) leads to the
non-diagonal Goldberger-Treiman relation $ C_5^A(Q^2)=f_\pi g_{\pi
N\Delta}(Q^2)/2M_N$ where $g_{\pi N\Delta}(Q^2)$ is determined
from the matrix element of the pseudoscalar density
 \be
 2m_q<\Delta^+ (p',s') | \bar{\psi} \gamma_5 \frac{\tau^3}{2} \psi |N(p,s) > =
 g_{\pi N\Delta}(Q^2)\sqrt{\frac{2}{3}}\frac{q_\sigma}{2M_N} \frac{m_\pi^2 f_\pi}{Q^2+m_\pi^2}\bar{u}_\sigma (p',s')
 u(p,s),
 \label{gpiND}
 \ee
where $m_q$ is the renormalized quark mass. The pion decay
constant, $f_\pi$, is determined from the two-point function $< O
| A^{a}_{\mu} | \pi^{b} (p) > = i p_\mu \delta^{ab} f_{\pi}$,
defined so that
 the continuum value is taken  $f_\pi=93.2$~MeV.
In order to relate the lattice pion matrix element to its physical
value we need the pseudoscalar renormalization constant, $Z_p$. We
take for quenched~\cite{renorm} and dynamical Wilson
fermions~\cite{renorm_dyn} $Z_p(\mu^2a^2\sim 1)=0.5(1)$ computed
using the RI-MOM method. This value may depend on the renormalization scale whereas it is not known
for domain wall fermions. In Fig.~\ref{fig:gpind} we show the
result for $g_{\pi N\Delta}$ for Wilson fermions and the linear
 extrapolation in $m_\pi^2$ of
these results to the chiral limit. We note that in the figure we
only show statistical errors which do not include a 10\%
uncertainty in $Z_p$. Furthermore, we would like to stress that
the determination of the quark mass $m_q$ using the axial Ward
identity has corrections of order $a$. These corrections become
more significant with decreasing quark mass. These can lead to
large uncertainties in the determination of $g_{\pi N\Delta}$.

In Fig.~\ref{fig:Goldberger-Treiman} we show the ratio $f_\pi
g_{\pi N\Delta}/\left(2 M_N C_5^A\right)$ for Wilson fermions. As
can be seen
this ratio is almost $Q^2$ independent 
and as the quark mass decreases it becomes consistent with unity
in agreement with the non-diagonal Goldberger-Treiman relation.

Under the assumptions that $C_3^A\sim 0$ and
that  $C_4^A$ is suppressed as compared to $C_5^A$,
both of which are justified by the lattice results, the parity
violation asymmetry can be shown to be proportional to the ratio
$C_5^A/C_3^V$~\cite{Nimai}. The form factor $C_3^V$ can be
obtained from the electromagnetic N to $\Delta$ transition. Using
our lattice results for the dipole and electric quadrupole Sachs
factors, ${\cal G}_{M1}$ and ${\cal G}_{E2}$~\cite{ND}, $C_3^V$ is
extracted from the relation $
C_3^V=\frac{3}{2}\frac{M_\Delta(M_N+M_\Delta)}{(M_N+M_\Delta)^2+Q^2}
\left({\cal G}_{M1}-{\cal G}_{E2}\right)$.

We show in Fig.~\ref{fig:CA5 over CV3}  the ratio $C_5^A/C_3^V$,
for pions of mass about 500 MeV.
As can be seen unquenching effects are small and in order
to assess the quark mass dependence we  extrapolate our quenched results, which
carry the smallest errors,
 to the chiral limit. We find only a small
increase in this ratio as we tune the quark mass to zero, indicating
a weak quark mass dependence. Therefore our lattice evaluation
 provides a prediction for the physical value of this ratio,
which is the analog of the $g_A/g_V$.
Our lattice results show  that this ratio, and therefore to
a first approximation the parity violating asymmetry, is non-zero
at $Q^2= 0$ and increases for $Q^2\stackrel{>}{\sim}1.5$~GeV$^2$.

\vspace*{-0.1cm}

\section{Conclusions}
\vspace*{-0.1cm}
In summary we have provided a lattice calculation
of the axial N to $\Delta$  transition form factors in the
quenched approximation, using two degenerate dynamical Wilson
fermions and within a hybrid approach where we use MILC
configurations and domain wall fermions.

The main conclusions are: 1.  $C_3^A$ is consistent with zero
whereas $C_4^A$ is small but shows the largest sensitivity to
unquenching effects. 2. The two dominant form factors are $C_5^A$
and $C_6^A$. These are  related in the chiral limit by axial
current conservation. The ratio $\left(Q^2/M_N^2\right)
C_6^A/C_5^A$, which must be unity if chiral symmetry is unbroken,
is shown to approach unity as the quark mass decreases. 3. For any
quark mass the strong coupling $g_{\pi N\Delta}$ and the axial
form factor $C_5^A$  show a similar $Q^2$ dependence with the
non-diagonal Goldberger-Treiman relation being reproduced as the
quark mass decreases. 4. The ratio of $C_5^A/C_3^V$ which
determines to a good approximation the parity violating asymmetry
is predicted to be non-zero at $Q^2=0$ and has a two-fold increase
when $Q^2\sim 1.5$~GeV.

\end{document}